\documentclass[a4paper, 12pt, onecolumn, nofootinbib,superscriptaddress]{revtex4}
\usepackage{amsfonts, amssymb, graphicx, float, appendix, multirow, longtable}
\usepackage{amsmath}
\usepackage{color}
\usepackage{hyperref}
\usepackage{mathrsfs}
\usepackage[left=2.5cm,top=2.0cm,right=2.5cm,bottom=3cm]{geometry}
\usepackage{subfigure}
\usepackage{xcolor}

\newcommand{\sfrac}[2]{{\textstyle{#1\over#2}}}

\newcommand{\dl}{\text{d}}

\begin{document}

\title{\textbf{Quasinormal modes of a Schwarzschild black hole within the Bondi-Sachs framework}}

\author{Bishop Mongwane}
\email{bishop.mongwane@uct.ac.za} 
\affiliation{Department of Mathematics \& Applied Mathematics,
  University of Cape Town, Rondebosch 7701, South Africa}

\author{Sipho Nkele}
\email{nklsip004@myuct.ac.za}
\affiliation{Department of Mathematics \& Applied Mathematics,
  University of Cape Town, Rondebosch 7701, South Africa}

\author{Didam G.A. Duniya}
\email{duniyaa@biust.ac.bw}
\affiliation{Department of Physics \& Astronomy,
  Botswana International University of Science and Technology, Palapye, Botswana}

\author{Nigel T. Bishop}
\email{n.bishop@ru.ac.za}
\affiliation{Department of Mathematics,
  Rhodes University, Grahamstown 6140, South Africa}

\begin{abstract}
Studies of quasinormal modes (QNMs) of black holes have a long and well established history. Predominantly, much research in this area has customarily focused on the equations given by Regge, Wheeler and Zerilli. In this work we study  linearized perturbations of a Schwarzschild black hole using the Characteristic formulation of numerical relativity, with an emphasis on the computation of QNMs. Within this formalism, the master equation describing gravitational perturbations is known to satisfy a fourth order differential equation. We analyse the singular points of this master equation, and obtain series solutions whose coefficients are given by three term recurrence relations, from which Leaver's continued fraction method can be applied. Using this technique, we recover the standard Schwarzschild quasinormal modes.
In addition, we find that imposing purely outgoing boundary conditions, a natural feature of the Bondi-Sachs framework, leads to the recovery of the algebraically special mode. 
\end{abstract}

\maketitle

\section{Introduction}

The field of black hole perturbation theory has evolved significantly since its inception, dating back to the seminal research of Regge and Wheeler \cite{regge:1957}. Their pioneering study explored non-spherical metric perturbations of Schwarzschild black holes and the linear stability of the Schwarzschild solution, heralding the advent of black hole perturbation theory. This groundbreaking work opened the door to a series of investigations throughout the 1970s, leading to crucial insights into the physics of perturbed spherical and rotating black holes. Other notable contributions during this golden age period include Zerilli's analysis of even-parity perturbations \cite{zerilli:1970}, Moncrief's gauge-invariant description of metric perturbations \cite{moncrief:1974}, Vishveshwara’s identification of quasinormal modes \cite{vishveshwara:1970}, and the subsequent analysis of these modes by Press \cite{press:1971} along with Chandrasekhar and Detweiler's numerical computations of the frequencies \cite{chandrasekhar:1975}. Subsequently, Teukolsky broadened this analysis to include the Kerr spacetime \cite{teukolsky:1972}. A comprehensive account of these developments can be found in the monograph by Chandrasekhar \cite{chandrasekhar:1983}.

A key component in studies of black hole perturbation theory is the computation of the quasinormal mode spectrum. Quasinormal modes (QNMs) represent the characteristic `ringing' of a black hole in response to external perturbations. These oscillations are characterized by complex frequencies that are unique to the mass, charge, and angular momentum of the black hole, thus making them a unique fingerprint of the black hole's properties. The importance of QNMs was highlighted by recent detections of gravitational waves from binary black hole mergers. As black holes are perturbed -- whether by matter falling into them, the merger of black holes in binary systems, or other astrophysical processes -- they emit gravitational waves that carry the signature of these quasinormal modes.  Performing black hole spectroscopy, especially with next-generation gravitational wave observatories like LIGO and LISA, will offer unprecedented opportunities for gravitational wave astronomy. Within the realm of numerical relativity, QNMs have been seen in simulations of binary neutron stars and black hole coalescence, where they are observed to dominate the radiated signal at intermediate times \cite{Anninos:1995vf,Shibata:2006nm}. These results are consistent with previously established knowledge based on perturbative calculations, which further highlights the synergy between numerical relativity and theoretical perturbative studies.

Aside from the algebraically special mode \cite{1984RSPSA.392....1C,couch:1973,Andersson_1994,1993GReGr..25.1185Q,MaassenvandenBrink:2000iwh}, it has not been possible, in general, to find explicit analytic expressions for the QNM eigenfrequencies in closed form. As a result, there are now a plethora of numerical and semi-analytical approximation techniques to compute black hole quasinormal modes that have emerged over the years. These can be broadly divided into methods that seek simple, approximate formulas that could, in principle, be more amenable to physical interpretation, e.g. \cite{1984PhRvD..30..295F,Dolan:2009nk} and methods that produce numerical results with high precision, e.g. \cite{1985RSPSA.402..285L,1993PhRvD..47.5253N,Saad:2003vhv,Cho:2009cj}. 
A survey of some of these methods can be found in the review articles \cite{1999CQGra..16R.159N,Kokkotas:1999bd,Berti:2009kk,Konoplya:2011qq}. A key feature in these techniques is that they are commonly applied to Regge-Wheeler, Zerilli and Teukolsky equations, or their equivalents in higher dimensions or other contexts. In this work, we obtain the master equation characterising gravitational perturbations from the Characteristic formulation.

The characteristic formulation of numerical relativity offers a powerful framework for analyzing black hole spacetimes. This formulation casts the gravitational field equations in terms of outgoing null hypersurfaces, facilitating the extraction of gravitational waves from numerically generated spacetimes. This extraction method is accomplished via the Cauchy Characteristic Extraction (CCE) algorithm where one takes metric data on some inner timelike worldtube $\Gamma$, computed from a 3+1 Cauchy code and propagate it to future null infinity $\mathscr{I}^{+}$ via a Characteristic code, thus enabling waveform extraction at $\mathscr{I}^{+}$ \cite{PhysRevD.54.6153,lrr:bishop}. This scheme represents a special case of the more ambitious Cauchy Characteristic Matching (CCM) \cite{1992anr..conf...20B,0264-9381-10-2-015} which, in turn, uses the data from the Characteristic code as exact boundary conditions for the metric functions of the 3+1 Cauchy code, consequently eliminating the need for outer boundary data in the Cauchy evolution.

Beyond gravitational wave extraction, the Bondi-Sachs framework has also found use in cosmological applications \cite{vanderWalt:2011jt}. Moreover, the formulation has, in recent times been applied to study perturbations of the Schwarzschild geometry in general relativity \cite{Bishop:2004ug,Bishop:2009ba,PhysRevD.87.104016,cedeno:2015ucy} and also in metric $f(R)$ gravity \cite{mongwane:2017}. In the context of QNMs, the formulation has been employed in \cite{Siebel:2001dp} to compute quasinormal modes of neutron stars, while in \cite{samuelsson:2007} Bondi-Sachs coordinates are used in the study of quasinormal modes of a toy problem, along with a constant density star. Both cases employ numerical integration and utilize grid compactification to impose boundary conditions at null infinity, facilitating a precise treatment of the asymptotic regime. In this work, we revisit the linearized perturbations of the Schwarzschild black hole using this formalism, further enriching our toolkit for exploring black hole quasinormal modes and the dynamics of black holes in general.

This paper is structured as follows: In \S\ref{sec:formalism} we review the formalism that is used to derive the master equation that governs gravitational perturbations within the characteristic formalism. Readers familiar with this topic may skip this section. In \S\ref{sec:maser_equation} we analyse the behaviour of the master equation at each of the singular points. This is followed by the derivation of the series solution and resulting continued fraction equation in \ref{eq:quasinormal_modes}. We cover the algebraically special mode in \S\ref{sec:algebraically_special_mode}. Some results are presented in \ref{sec:numerical_results} and finally, we conclude in \S\ref{sec:concluding_remarks}. Throughout this paper, we use Geometrized units $G=c=1$ and metric signature $(-+++)$. Moreover, in our equations, we use $r$, the Schwarzschild radial coordinate and not the \textit{tortoise coordinate} that is often used in discussions of black hole perturbations.

\section{Formalism}
\label{sec:formalism}

\subsection{Bondi-Sachs metric}
The Bondi-Sachs formalism~\cite{Bondi21,Sachs103} was developed in 1962 and established the energy transported to infinity by gravitational waves. It is also discussed in several recent reviews~\cite{lrr:bishop,lrr:winicour,Madler:2016xju}.
Using coordinates $x^a=(u,r,x^A)$ with $a=0,\cdots,3$ and $A=2,3$, the Bondi-Sachs metric is
\begin{align}
  \label{eq:bs_metric}
ds^{2} =&-\left(e^{2\beta}\left(1+\frac{W}{r}\right)-r^{2}h_{AB}U^{A}U^{B} \right)\dl u^{2}-2e^{2\beta}\dl u \dl r \\\nonumber
&-2r^{2}h_{AB}U^{B}\dl u \dl x^{A}+r^{2}h_{AB}\dl x^{A}\dl x^{B} \;.
\end{align}
The lines $x^A,u=$ constant are outgoing light-rays, and $r$ is a surface area coordinate meaning that the 2-surfaces $u,r=$ constant have area $4\pi r^2$. The $x^A$ can be general angular coordinates but are taken here to be spherical polars $\theta,\phi$. The surface area condition on $r$ implies that $\det(h_{AB})=\det(q_{AB})$ where $q_{AB}$ is the spherically symmetric 2-metric $q_{AB}\dl x^A\dl x^B=\dl\theta^2+\sin^2\theta\dl\phi^2$. The Einstein equations are much simplified~\cite{bishop-hpgn1997,gomez1997,lrr:bishop} on the introduction of a complex dyad  $q^A=(1,i/\sin\theta),q_A=(1,i\sin\theta)$, and then representing $U^A,h_{AB}$ by the spin-weighted quantities
\begin{equation}
U=U^A q_A\,,\;J=\frac{h_{AB}q^Aq^B}{2}\,.
\end{equation}
The normalization on $q^A$ is $q^Aq^Bq_{AB}=2$, which differs from the normalization to $1$ and denoted as $m^A$ used originally~\cite{NewmanPenrose1963}. Note that, as defined above and in a gauge in which all metric coefficients fall off as $r^{-1}$, $J=h_+ +ih_\times$ where $h_+,h_\times$ are the standard polarization components of a gravitational wave in the TT gauge. Angular derivatives are defined in terms of the complex differential operators $\eth,\bar{\eth}$
\begin{equation}
\eth f=q^A\partial_A f - s f \cot\theta\,,\;\;
\bar{\eth} f=\bar{q}^A\partial_A f + s f \cot\theta\,,
\end{equation}
where $s$ is the spin-weight of $f$: $s=0$ for $\beta$ and $W$, $s=1$ for $U$, and $s=2$ for $J$~\cite{gomez1997,lrr:bishop}.

\subsection{Linearized field equations}
The Schwarzschild geometry can be represented in Bondi-Sachs form by the Eddington-Finkelstein metric
\begin{align}
\label{eq:background}
ds^{2} = - \left(1-\frac{2M}{r} \right)\dl u^{2}-2\dl u\dl r + r^{2}q_{AB}\dl x^{A}\dl x^{B}\;,
\end{align}
and we then consider small perturbations about this background by making the ``separation of variables'' ansatz
\begin{align}
\beta&=\Re(\beta_0(r) e^{\rho u})Z_{\ell,m}\,,\;W=-2M+w\mbox{ where }w=\Re(w_0(r) e^{\rho u})Z_{\ell,m}\,,\;\nonumber\\
U&=\Re(U_0(r) e^{\rho u})\eth Z_{\ell,m}\,,\;J=\Re(J_0(r) e^{\rho u})\eth^2 Z_{\ell,m}\,.
\label{e-ansatz}
\end{align}
The $Z_{\ell,m}$ are real functions of $x^A$ constructed from the usual spherical harmonic functions $Y_{\ell,m}$~\cite{Zlochower03,lrr:bishop}, but note that $\eth Z_{\ell,m},\eth^2 Z_{\ell,m}$ are in general complex. The quantity $\rho$ is complex: $\Im(\rho)$ is the wave frequency, and $\Re(\rho)\le 0$, with $\Re(\rho)< 0$ meaning that the wave is damped. A general perturbation is obtained by summing the right hand side of Eq.~\eqref{e-ansatz} over $\ell,m$ and integrating over $\rho$, but that is not needed here as the QNM calculations regard the perturbation as comprising a single mode: thus, $\ell,m$ and $\rho$ are treated as constant.

The linearized Einstein equations are obtained by treating $\beta,w,U,J$ as small quantities with any product ignorable. The equations that will be needed are:
\begin{subequations}
\begin{eqnarray}
R_{11}&:\; &\frac{4}{r} \beta_{,r}=0,
\label{e-b}\\
q^{^{_A}} R_{1A}&:\; &\frac{1}{2r} \left(
4 \eth \beta - 2 r \eth \beta_{,r} + r \bar{\eth} J_{,r}
+r^3  U_{,rr} +4 r^2 U_{,r} \right) = 0,
\label{e-rq} \\
q^{^{_A}} q^{^{_{_{B}}}} R_{_{AB}}&:\;
  &2 r (rJ)_{,ur}-2\eth^2\beta + (r^2 \eth U)_{,r} - 2(r-M) J_{,r}
 -  r^2J_{,rr}\left(1-\frac{2M}{r}\right)   
  = 0.
\label{e-ev}
\end{eqnarray}
\end{subequations}
Eq.\eqref{e-b} implies $\beta(r)=\beta_0=$constant, and then Eqs.~\eqref{e-ansatz}, \eqref{e-rq} and \eqref{e-ev} give
\begin{subequations}
\label{eq:linear_eqs} 
\begin{eqnarray}
\label{eq:lU}
q^AR_{1A}&:&r^{3}U_{0,rr}+4r^{2}U_{0,r}+[2-\ell(\ell+1)]rJ_{0,r}+4\beta_0 - 2r\beta_{0,r} =0,\\
q^Aq^BR_{AB}&:&2r\rho\left(rJ_0 \right)_{,r} -2\beta_0 +(r^{2}U_{0})_{,r}-2(r-M)J_{0,r}
-r^{2}J_{0,rr}\left(1-\frac{2M}{r} \right)= 0 ,
\label{eq:lJ}
\end{eqnarray}
\end{subequations}
where evaluation of $\bar{\eth}J$ has used $\bar{\eth}\eth^2 Z_{\ell,m}=(2-\ell(\ell+1))\eth Z_{\ell,m}$. Then Eqs. \eqref{eq:lU} and \eqref{eq:lJ} are solved simultaneously, as described in \S\ref{sec:maser_equation} to find $U_0,J_0$ and thus $J,U$.
  
While it is not needed here, the complete metric can now be obtained~\cite{Bishop:2004ug}: $U^2=\Re(U),U^3=\Im(U)/\sin\theta$, and $h_{AB}$ is given explicitly in Eq.~(424) of~\cite {lrr:bishop}; $R_{AB}h^{AB}=0$ reduces to $w_{0,r}=f(U_0,J_0,\beta_0)$ so leading to $w_0$; and $R_{00}=0,R_{01}=0,q^AR_{0A}=0$ are constraints which can be used to eliminate some constants of integration.

\section{Master equation}
\label{sec:maser_equation}

The equations \eqref{eq:lU}, \eqref{eq:lJ} and $\partial_r$\eqref{eq:lJ} constitute a system of 3 linear algebraic equations in the 3 variables $U_0,U_{0,r},U_{0,rr}$ and are solved using standard techniques. The result is that we obtain expressions for $U_0,U_{0,r},U_{0,rr}$ in terms of $J_0$ and its derivatives. Then evaluation of the identity $\partial_r U_0-U_{0,r}=0$ leads to
\begin{eqnarray}
  \label{eq:master_eq_fourth_order}
  &&r^3(r-2M)J_{,rrrr}-2r^2\left(\rho r^2-4r+5M\right)J_{,rrr} -r \left[14\rho r^2-14r +\ell(\ell+1) r+4M\right]J_{,rr}\nonumber \\
  &&-2\left[8\rho r^2+\ell(\ell+1) r -2r-2M \right]J_{,r} = 0\;.
\end{eqnarray}
This master equation was first derived in \cite{Bishop:2004ug}, for the $\ell=2$ case, and in terms of the variable $x=1/r$. Being a fourth order ODE, a consistent analysis of Eq. (\ref{eq:master_eq_fourth_order}) in the current form can be a fiendishly challenging task. It can, however, be simplified via a judicious choice of variable. By making the transformation $\mathcal{J}=r^3(rJ(r))_{,rr}$, Eq. (\ref{eq:master_eq_fourth_order}) reduces to
\begin{equation}
  \label{eq:master_eq}
  r^2\left(r-2M \right)\mathcal{J}_{,rr}-2r\left(\rho r^2+r-5M \right)\mathcal{J}_{,r}-\left[2\rho r^2+(\ell^2+\ell-2)r+16M \right]\mathcal{J} = 0\;.
\end{equation}
We can now turn our attention to the singular points of \eqref{eq:master_eq}.
\begin{itemize}
\item \textit{Regular singular point at $r=0$}
  \\If we express the solutions of the master equation (\ref{eq:master_eq}) near the singularity at $r=0$, as
  \begin{equation}
    \lim_{r\rightarrow 0} \mathcal{J} = r^{k_i}\;,
  \end{equation}
  then the exponents $k_i$ take the values $k_1=4$ and $k_2= 2$. The point at the origin is rarely considered in the context of quasinormal modes studies. However it can be useful when studying the algebraically special mode \cite{MaassenvandenBrink:2000iwh}. We will return to this point in \S\ref{sec:algebraically_special_mode}. In the meantime, it is worth pointing out that one can obtain two linearly independent solutions about the point $r=0$. Moreover, since $k_1-k_2\in\mathbb{Z}$ one of the solutions can be obtained by the method of Fr\"{o}benius, while the second solution contains a logarithmic term, that may vanish under certain conditions.
  
\item \textit{Regular singular point at $r=2M$}
  \\On the other hand, if we express the solutions of the master equation (\ref{eq:master_eq}) near the singularity at $r=2M$, as
  \begin{equation}
    \lim_{r\rightarrow 2M} \mathcal{J} = (r-2M)^{s_i}\;,
  \end{equation}
  then the exponents $s_i$ take the values $s_1=0$ and $s_2= 4\rho M-2$. In the sign conventions adopted here, we will interpret $s_2$ as corresponding to the outgoing (from the horizon) solution.
\item \textit{Irregular singular point at $r=\infty$}
  \\Finally, the asymptotic solutions of the master equation (\ref{eq:master_eq}) at infinity are
  \begin{equation}
    \mathcal{J} = \sfrac{1}{r} \qquad \text{and} \qquad \mathcal{J} = r^{3+4\rho M}e^{2\rho r}\;.
  \end{equation}
  We note that the frequency $\nu$ (with $\rho=i \nu$) is complex, and its imaginary part governs the growth or decay of the perturbation field $\mathcal{J}$. If $\nu$ has a negative imaginary part, the modes will decay and the black hole is stable under the perturbation. We therefore expect the solution $\mathcal{J} = r^{3+4\rho M}e^{2\rho r}$ to serve as the outgoing mode.

\end{itemize}

\subsection{Quasinormal modes}
\label{eq:quasinormal_modes}
Black hole quasinormal modes are traditionally described as those solutions of the master equation having purely ingoing boundary conditions at the horizon and purely outgoing wave conditions at infinity. In the current context, we maintain the requirement of purely outgoing conditions at infinity, but for the boundary at the horizon $r=2M$, we simply require the solutions to not violate the causal condition that no information can leak from within the horizon. We therefore need to consider the master equation (\ref{eq:master_eq}) as a boundary value problem subject to the boundary conditions
\begin{equation}
  \mathcal{J} \xrightarrow{r\rightarrow 2M} (r-2M)^{0} \qquad \text{and} \qquad  \mathcal{J} \xrightarrow{r\rightarrow \infty} r^{4\rho M+3}\,e^{2\rho \,r}\;.
\end{equation}
This behaviour can be achieved by seeking power series solutions of the form
\begin{equation}
  \label{eq:series0}
  \mathcal{J} =  r^{4\rho M+3}\,e^{2\rho(r-2M)} \sum_{n=0}^{\infty}a_n\left(\frac{r-2M}{r} \right)^n\;.
\end{equation}
Substituting this series into the master equation (\ref{eq:master_eq}) leads to the three term recurrence relation that determines the expansion coefficients $a_n$
\begin{subequations}
  \label{eq:reccurence_qnm}
\begin{eqnarray}
  \alpha_0a_1+\beta_0a_0 &=& 0\;,\\
  \alpha_na_{n+1}+\beta_na_n+\gamma_na_{n-1}&=&0\qquad n=1,2,\cdots
\end{eqnarray}
\end{subequations}
where
\begin{eqnarray}
  \alpha_n &=& -n^2+(4M\rho-4)n+4M\rho-3\;,\\
  \beta_n &=& 2n^2-(16M\rho-2)n+32M^2\rho^2-8M\rho-3+\ell(\ell+1)\;,\\
  \gamma_n &=&-n^2+(8M\rho+2)n-16M^2\rho^2-8M\rho \;.
\end{eqnarray}
Three term recurrence relations such as Eq (\ref{eq:reccurence_qnm}) have been studied extensively in the literature \cite{gautschi:1967}. Typically, such relations admit two linearly independent solution sequences $f_n$ and $g_n$, characterized by the asymptotic property that $\lim_{n\rightarrow \infty} f_n/g_n=0$. In this case $f_n$ is referred to as the minimal solution, while $g_n$ is identified as the dominant solution.

The convergence of (\ref{eq:series0}) may be analysed by considering the large $n$ behaviour of the expansion coefficients $a_n$. Following \cite{baber:1935,1985RSPSA.402..285L} we expand $a_{n+1}/a_n$ in inverse fractional powers of $n$. The resulting expression
\begin{equation}
  \label{eq:asymptotic_0}
  \lim_{n\rightarrow \infty} \frac{a_{n+1}}{a_n} = 1 \pm \frac{\sqrt{-4M\rho}}{\sqrt{n}}-\frac{4M\rho+11/4}{n}
\end{equation}
lends itself to effortless application of the ratio test. In addition, we can deduce the large $n$ behaviour of $a_n$ from (\ref{eq:asymptotic_0}) via integration
\begin{equation}
  \label{eq:asymptotic_1}
  a_n \xrightarrow{n\rightarrow \infty} n^{-4M\rho-11/4}e^{\pm \sqrt{-16M\rho n}}\;.
\end{equation}
The two signs $\pm$ in (\ref{eq:asymptotic_0}) and (\ref{eq:asymptotic_1}) indicate the asymptotic behavior of the two independent solution sequences to the recurrence relation, with the ($-$) sign corresponding to the minimal solution and the ($+$) sign to the dominant solution. It follows that the series (\ref{eq:series0}) will converge if we choose the solution of (\ref{eq:reccurence_qnm}) which behaves asymptotically like $ a_n \rightarrow n^{-4M\rho-11/4}e^{- \sqrt{-16M\rho n}}$,  corresponding to the minimal solution sequence.

The process of obtaining the required minimal solution is known \cite{gautschi:1967,1985RSPSA.402..285L}. Using the same procedure as in \cite{1985RSPSA.402..285L}, the recurrence relations above can be used to derive an infinite continued fraction equation for the QNM eigenfrequencies
\begin{equation}
0 = \beta_0 - \cfrac{\alpha_0 \gamma_1}{\beta_1 - \cfrac{\alpha_1 \gamma_2}{\beta_2 - \cfrac{\alpha_2 \gamma_3}{\beta_3 - \cdots}}}
\end{equation}
It is notationally convenient to use the following form for the continued fraction above
\begin{equation}
  \label{eq:continued_fraction0}
  0 = \beta_0-\frac{\alpha_0 \gamma_1}{\beta_1-}\,\frac{\alpha_1 \gamma_2}{\beta_2-}\,\frac{\alpha_2 \gamma_3}{\beta_3-}\cdots
\end{equation}
Furthermore, as pointed out in \cite{1985RSPSA.402..285L}, this equation can be inverted an arbitrary number of times, with the n$th$ inversion given by
\begin{equation}
  \label{eq:continued_fraction1}
  \left[\beta_n-\frac{\alpha_{n-1}\gamma_n}{\beta_{n-1}-}\,\frac{\alpha_{n-2}\gamma_{n-1}}{\beta_{n-2}-}\,\cdots\, \frac{\alpha_0\gamma_1}{-\beta_0}\right] = \frac{\alpha_{n} \gamma_{n+1}}{\beta_{n+1}-}\,\frac{\alpha_{n+1} \gamma_{n+2}}{\beta_{n+2}-}\,\frac{\alpha_{n+2} \gamma_{n+3}}{\beta_{n+3}-}\cdots
\end{equation}
Both (\ref{eq:continued_fraction0}) and (\ref{eq:continued_fraction1}) are equivalent (for $n>0$) in that a root of (\ref{eq:continued_fraction1}) is also a root of (\ref{eq:continued_fraction0}), and vice-versa. However, for practical computations, we use (\ref{eq:continued_fraction1}). In this form, we compute the $n$th overtone as the root of the $n$th inversion.
%
\subsection{Algebraically special mode}
\label{sec:algebraically_special_mode}
As highlighted in \cite{1984RSPSA.392....1C}, algebraically special perturbations are those that excite gravitational waves that are either purely ingoing or purely outgoing. The Bondi-Sachs formalism is well suited to tackle these kinds of modes. In the following, we consider the purely outgoing modes in the master equation \ref{eq:master_eq}. In this case, we impose the following boundary conditions
\begin{equation}
  \mathcal{J} \xrightarrow{r\rightarrow 2M} (r-2M)^{4\rho M-2} \qquad \text{and} \qquad  \mathcal{J} \xrightarrow{r\rightarrow \infty} r^{4\rho M+3}\,e^{2\rho \,r}\;.
\end{equation}
The appropriate series, satisfying these conditions at the boundary, can be constructed in the form
\begin{equation}
  \label{eq:series1}
  \mathcal{J} = (r-2M)^{4\rho M-2} r^5\,e^{2\rho(r-2M)} \sum_{n=0}^{\infty}a_n\left(\frac{r-2M}{r} \right)^n\;.
\end{equation}
Now, after substituting (\ref{eq:series1}) into the master equation (\ref{eq:master_eq}), the sequence of expansion coefficients $a_n$ is determined by the three term recurrence relation
\begin{subequations}
  \label{eq:rercurrence_relations0}
  \begin{eqnarray}
    \label{eq:purely_outogoing_reccurrence_relation}
  \alpha_0a_1+\beta_0a_0 &=& 0\;,\\
  \alpha_na_{n+1}+\beta_na_n+\gamma_na_{n-1}&=&0\qquad n=1,2,\cdots
\end{eqnarray}
\end{subequations}
with
\begin{eqnarray}
  \alpha_n &=& -(n+1)(n-1+4\rho M)\;, \\
  \beta_n &=& 2n^2-6n+1+\ell(\ell+1)\;, \\
  \gamma_n &=& -(n-2)(n-4)\;.
\end{eqnarray}
The root equation determining $\nu$ takes the same form as (\ref{eq:continued_fraction1}).
\begin{table}[!h]
  \centering
  \begin{tabular}{crr}
    \hline
    \hline
    $\ell$ &  $2M\nu$ $(n=1)$ & $2M\nu$ $(n=3)$\\
    \hline
    \hline
    $2$ & $3.56\times10^{-25}-4i$&$1.69\times10^{-35}+4i$ \\
    $3$ & $-1.55\times10^{-23}-20i$& $-2.79\times10^{-33}+20i$ \\
    $4$ & $-4.20\times10^{-23}-60i$& $6.23\times10^{-48}+60i$\\
    $5$ & $-2.06\times10^{-22}-140i$& $1.00\times10^{-38}+140i$ \\
    $6$ & $-6.41\times10^{-23}-280i$& $-5.47\times10^{-32}+280i$\\
    \hline
  \end{tabular}
  \caption{Eigenfrequencies determined from Eqs (\ref{eq:purely_outogoing_reccurrence_relation}), for the cases $n=1$ and $n=3$.}
  \label{tab:algebraically_special}
\end{table}
Numerical results for $n=1$ and $n=3$ are given in Table \ref{tab:algebraically_special}. There are other roots corresponding to other $n$ values. Nevertheless, the roots corresponding to $n=1$ and $n=3$ are noteworthy; they converge to a high degree of accuracy, suggesting that they may in fact be exact solutions. This is indeed the case, as can be verified as follows. One can impose conditions under which the infinite series (\ref{eq:series1}) terminates, resulting in a finite number of equations for the $a_i$ coefficients, see e.g. \cite{1996CQGra..13..233L}. From the structure of the recurrence relations (\ref{eq:rercurrence_relations0}), this will be so if both $\gamma_{n+1}=0$ and $a_{n+1}=0$ for some $n$. If this holds, then $a_{n}=0$ for all $n>m$. Now, solving $\gamma_{n+1}=0$ leads to
\begin{equation}
  \gamma_{n+1} = -(n-1)(n-3) = 0\;,
\end{equation}
with the solutions $n=1$ and $n=3$. If in addition we impose $a_{n+1}=0$, we have the following cases
\begin{itemize}
\item Case $n=1$
  \\Imposing the condition $a_2=0$ in (\ref{eq:rercurrence_relations0}) results in
  \begin{subequations}
  \label{eq:runcated_relations_1}  
  \begin{eqnarray}
    \alpha_0a_1+\beta_0a_0 &=& 0\;,\\
    \beta_1a_1+\gamma_1a_{0}&=&0\;.
  \end{eqnarray}
  \end{subequations}
  This (\ref{eq:runcated_relations_1}) can be simplified to
  \begin{equation}
    \alpha_0\gamma_1 = \beta_0\beta_1\;,
  \end{equation}
  which is equivalent to the solution
  \begin{equation}
    \rho = \frac{(\ell-1)\ell(\ell+1)(\ell+2)}{12M}\;.
  \end{equation}
\item Case $n=3$
  \\Imposing $a_{n+1}=0$ leads to $a_4=0$, and the truncated recurrence relations become
  \begin{subequations}
    \begin{eqnarray}
      \label{eq:se1}
      \alpha_0a_1+\beta_0a_{0} &=& 0 \;,\\
      \label{eq:se2}
      \alpha_1a_2+\beta_1a_{1} +\gamma_{1}a_0&=& 0 \;,\\
      \label{eq:se3}
      \alpha_2a_3+\beta_2a_{2} +\gamma_{2}a_1&=& 0 \;,\\
      \label{eq:se4}
      \beta_3a_{3} +\gamma_{3}a_2&=& 0 \;.
  \end{eqnarray}
  \end{subequations}
  In this case, the coefficients $a_i$ $(i=1,2,3)$ can be computed explicitly (from Eqs (\ref{eq:se1})--(\ref{eq:se3}))
  \begin{eqnarray}
    a_1 &=& -\frac{\beta_0}{\alpha_0}a_0 \;,\\
    a_2 &=& \left[\frac{\beta_1}{\alpha_1}\frac{\beta_0}{\alpha_0} + \frac{\gamma_1}{\alpha_1}\right]a_0 \;,\\
    a_3 &=& \left[-\frac{\beta_2}{\alpha_2}\frac{\beta_1}{\alpha_1}\frac{\beta_0}{\alpha_0} - \frac{\beta_2}{\alpha_2}\frac{\gamma_1}{\alpha_1} + \frac{\gamma_2}{\alpha_2}\frac{\beta_0}{\alpha_0} \right]a_0\;.
  \end{eqnarray}
  One can then solve (\ref{eq:se4})
  \begin{equation}
    \left[-\frac{\beta_2}{\alpha_2}\frac{\beta_1}{\alpha_1}\frac{\beta_0}{\alpha_0} - \frac{\beta_2}{\alpha_2}\frac{\gamma_1}{\alpha_1} + \frac{\gamma_2}{\alpha_2}\frac{\beta_0}{\alpha_0} \right] \beta_3 +\left[\frac{\beta_1}{\alpha_1}\frac{\beta_0}{\alpha_0} + \frac{\gamma_1}{\alpha_1}\right]\gamma_{3}= 0\;,
  \end{equation}
  which yields
  \begin{equation}
    \label{eq:asmm}
    \rho =- \frac{(\ell-1)\ell(\ell+1)(\ell+2)}{12M}\;.
  \end{equation}
\end{itemize}
Interestingly, the root (\ref{eq:asmm}) arises at the origin as follows. A series solution about the regular singular point $r=0$ can be constructed by the Fr\"{o}benius method,
\begin{equation}
  \mathcal{J}_1 = c_0 r^4 \left(1-\frac{(\ell-2)(\ell+3)}{6M}r - \frac{60\rho M - (\ell-2)(\ell-3)(\ell+3)(\ell+4)}{96M^2}r^2 +\cdots\right)\;.
\end{equation}
Since the roots of the indicial equation differ by an integer, the second solution has the form
\begin{equation}
  \label{eq:second_solution}
  \mathcal{J}_2 = C\,\mathcal{J}_1\,\ln(r)  + r^{2}\sum_{i=0}^{\infty}d_jr^{j}\;,
\end{equation}
where $C$ is given by
\begin{equation}
  C = -\frac{(\ell-1)(\ell)(\ell+1)(\ell+2) +12\rho M}{8M^2}\;.
\end{equation}
Note that the second solution (\ref{eq:second_solution}) diverges as $r\rightarrow 0$. However, a \textit{miraculous} cancellation ($C=0$) of this term occurs when $\rho$ is given by equation (\ref{eq:asmm}) in which case the second solution becomes a power series, see also \cite{MaassenvandenBrink:2000iwh}.

\section{Numerical results}
\label{sec:numerical_results}

The continued fraction equations (\ref{eq:continued_fraction1}) is computed via Lentz' algorithm \cite{lentz:1976,thompson:1986}. This presents us with a standard root finding problem for each multipole number $\ell$. Moreover, for each $\ell$, the continued fraction equation has an infinite number of roots which are characterized by the overtone number $n$. The root finding methods employed here require initial guesses for each $n$. When $n>2$, we set the guess for $\rho$ via linear extrapolation from the previous roots. For $n=1$ and $n=2$, we initially do not have any data to extrapolate from. For that case, the initial guess is obtained via an expansion for the frequency in inverse powers of $L=\ell+1/2$. In the notation of \cite{Dolan:2009nk}, this gives $\rho$ as
\begin{equation}
-i  \rho_{\ell n} = \varpi^{(n)}_{-1}L+\varpi^{(n)}_{0}+\varpi^{(n)}_{1}L^{-1}+\varpi^{(n)}_{2}L^{-2}+\cdots
\end{equation}
The expressions of the $\varpi_{i}^{(n)}$ are given in \cite{Dolan:2009nk}. In principle, one could do away with the extrapolation step by using an asymptotic expansion for the modes $\rho_{\ell n}$ as initial guesses. For the Schwarszchild case, the asymptotic expression is known to be \cite{1984PhRvD..30..295F}
\begin{equation}
  -i  \rho_{\ell n} = \frac{\ell+\frac{1}{2}-(n+\sfrac{1}{2})i}{\sqrt{27}M}\;.
\end{equation}
However, since we are interested in further applying this technique to other scenarios where simple asymptotic expansions may not be readily available, we restrict our approach here to using linear extrapolation for the initial guesses.

\begin{figure}[!h]
  \centering
  \includegraphics[width=0.75\textwidth]{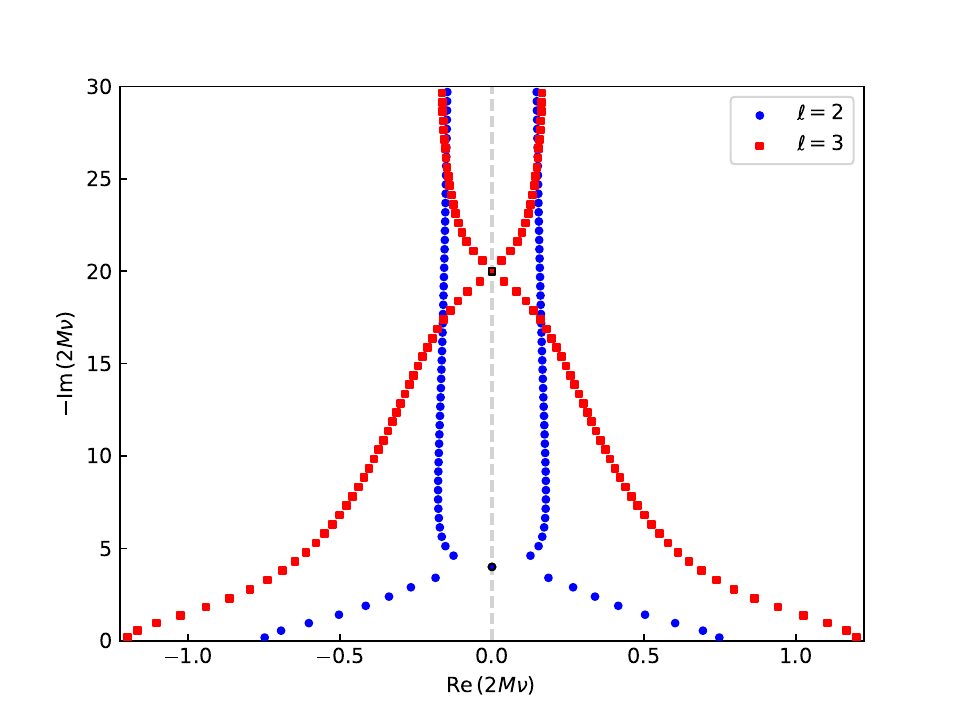}
  \caption{The first $60$ Schwarzschild quasinormal frequencies for $\ell=2$ and $\ell=3$. The algebraically special modes have black outlines over them.}
  \label{fig:l2_l3}
\end{figure}

\begin{figure}[!h]
  \centering
  \includegraphics[width=0.75\textwidth]{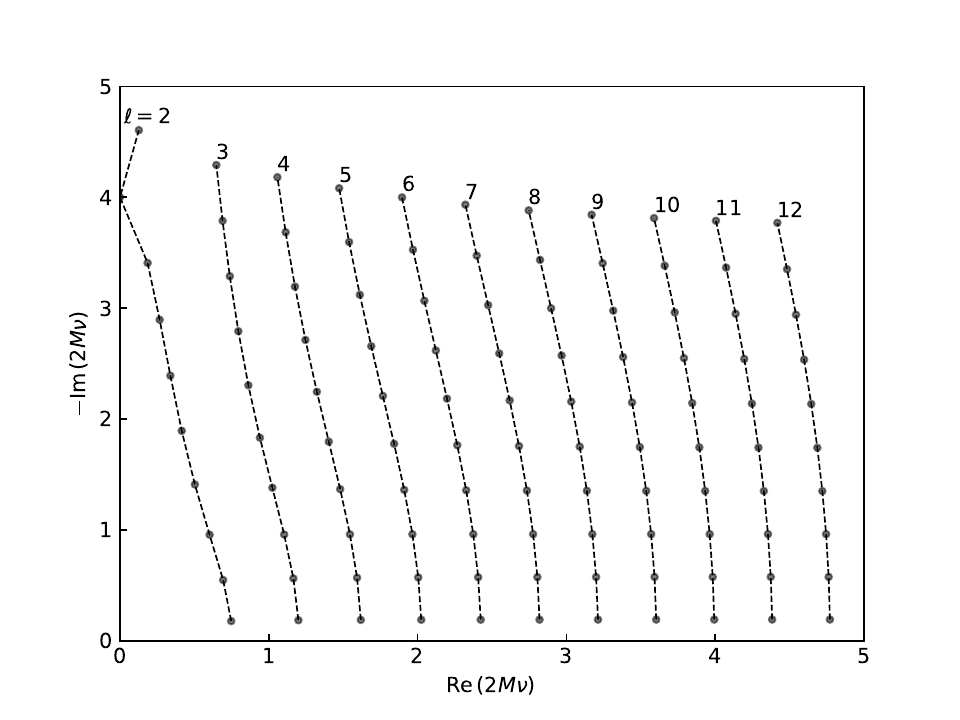}
  \caption{First $10$ Schwarzschild quasinormal frequencies for $\ell=2$ through $\ell=12$. The dashed lines connect different overtones of the modes corresponding to the same value of $\ell$}
  \label{fig:l2_l12}
\end{figure}

In Fig \ref{fig:l2_l3}, we plot the first $60$ modes for $\ell=2$ and $\ell=3$. In Fig \ref{fig:l2_l12} we plot the first $10$ modes for $\ell=2,3,\cdots,12$. In each case, the results are consistent with the standard quasinormal modes of the Schwarszchild black hole.

\section{Concluding remarks}
\label{sec:concluding_remarks}

In this study, we have successfully shown that it is feasible to compute the quasinormal modes of black holes using the Characteristic formulation framework, thus filling a previously missing component in Bondi-Sachs perturbation analyses of black holes. This development not only serves as a testament to the versatility of the Characteristic formulation but also as a bridge connecting theoretical insights with practical numerical relativity methods. Such a bridge is crucial if one is to take advantage of the latest advances in gravitational wave astronomy.

There are a number of possible extensions of the work presented here. A natural extension is the study of perturbations of spherically symmetric systems in $f(R)$ gravity. Notably, it has been established that tensor mode perturbations of a Schwarzschild black hole retain their form as in general relativity when extended to $f(R)$ gravity, e.g. \cite{mongwane:2017,nzioki:2017}. In this scenario, the main new features arise from scalar mode perturbations, whose master equation takes the form of a massive Klein-Gordon field equation. Another avenue is the study of linearized charged black holes, where the solutions to the master equation might extend beyond traditional three-term recurrence relations. Overall, we envisage that this approach to perturbation theory is widely applicable to a variety of astrophysical situations, including perturbations of the Kerr black hole, whose metric in Bondi-Sachs form is presented in \cite{nigel:2006}, see also \cite{bai:2007,arganaraz:2021,maedler:2023} for alternative representations of the Kerr spacetime in characteristic coordinates. Lastly, we note that the Schwarzschild black hole, in general relativity, is stable as has been rigorously shown in the literature \cite{vishveshwara:1970,wald:1979,kay:1987}. This stability is reflected in the negative imaginary part of the quasi-normal mode frequencies, which ensures the exponential decay of perturbations over time. In contrast, instabilities may arise in some of the cases outlined above \cite{Konoplya:2011qq}.


\begin{thebibliography}{10}

\bibitem{Andersson_1994}
N.~Andersson.
\newblock Total transmission through the schwarzschild black-hole potential
  barrier.
\newblock {\em Classical and Quantum Gravity}, 11(3):L39, mar 1994.

\bibitem{Anninos:1995vf}
P.~Anninos, R.~H. Price, J.~Pullin, E.~Seidel, and W.-M. Suen.
\newblock {Headon collision of two black holes: Comparison of different
  approaches}.
\newblock {\em Phys. Rev. D}, 52:4462--4480, 1995 arXiv:gr-qc/9505042.

\bibitem{Berti:2009kk}
E.~Berti, V.~Cardoso, and A.~O. Starinets.
\newblock {Quasinormal modes of black holes and black branes}.
\newblock {\em Class. Quant. Grav.}, 26:163001, 2009 arXiv:0905.2975,  [gr-qc].

\bibitem{1992anr..conf...20B}
N.~T. {Bishop}.
\newblock {Some aspects of the characteristic initial value problem in
  numerical relativity.} In {\em Approaches to Numerical Relativity}.
\newblock R.~{D'Inverno}, editor, pages 20--33, 1992.

\bibitem{0264-9381-10-2-015}
N.~T. Bishop.
\newblock Numerical relativity: combining the cauchy and characteristic initial
  value problems.
\newblock {\em Classical and Quantum Gravity}, 10(2):333, 1993.

\bibitem{Bishop:2004ug}
N.~T. Bishop.
\newblock {Linearized solutions of the Einstein equations within a Bondi-Sachs
  framework, and implications for boundary conditions in numerical
  simulations}.
\newblock {\em Class. Quant. Grav.}, 22:2393--2406, 2005 arXiv:gr-qc/0412006,
  [gr-qc].

\bibitem{PhysRevD.54.6153}
N.~T. Bishop, R.~G\'omez, L.~Lehner, and J.~Winicour.
\newblock Cauchy-characteristic extraction in numerical relativity.
\newblock {\em Phys. Rev. D}, 54:6153--6165, Nov 1996.

\bibitem{bishop-hpgn1997}
N.~T. Bishop, R.~G\'omez, L.~Lehner, and J.~Winicour.
\newblock High-powered gravitational news.
\newblock {\em Phys. Rev. D}, 56:6298--6309, Nov 1997.


\bibitem{Bishop:2009ba}
N.~T. Bishop and A.~S. Kubeka.
\newblock {Quasi-Normal Modes of a Schwarzschild White Hole}.
\newblock {\em Phys. Rev. D}, 80:064011, 2009 arXiv:0907.1882,  [gr-qc].

\bibitem{lrr:bishop}
N.~T. Bishop and L.~Rezzolla.
\newblock Extraction of gravitational waves in numerical relativity.
\newblock {\em Living Reviews in Relativity}, 20(1):1, 2016.

\bibitem{nigel:2006}
N.~T. Bishop and L.~R. Venter.
\newblock Kerr metric in bondi-sachs form.
\newblock {\em Phys. Rev. D}, 73:084023, Apr 2006.

\bibitem{Bondi21}
H.~Bondi, M.~G.~J. van~der Burg, and A.~W.~K. Metzner.
\newblock Gravitational waves in general relativity. vii. waves from
  axi-symmetric isolated systems.
\newblock {\em Proceedings of the Royal Society of London A: Mathematical,
  Physical and Engineering Sciences}, 269(1336):21--52,
  1962http://rspa.royalsocietypublishing.org/content/269/1336/21.full.pdf.

\bibitem{chandrasekhar:1983}
S.~Chandrasekhar.
\newblock {\em The Mathematical Theory of Black Holes}.
\newblock International series of monographs on physics. Clarendon Press, 1983.

\bibitem{1984RSPSA.392....1C}
S.~{Chandrasekhar}.
\newblock {On Algebraically Special Perturbations of Black Holes}.
\newblock {\em Proceedings of the Royal Society of London Series A},
  392(1802):1--13, March 1984.

\bibitem{chandrasekhar:1975}
S.~{Chandrasekhar} and S.~{Detweiler}.
\newblock {The Quasi-Normal Modes of the Schwarzschild Black Hole}.
\newblock {\em Proceedings of the Royal Society of London Series A},
  344(1639):441--452, August 1975.

\bibitem{Cho:2009cj}
H.~T. Cho, A.~S. Cornell, J.~Doukas, and W.~Naylor.
\newblock {Black hole quasinormal modes using the asymptotic iteration method}.
\newblock {\em Class. Quant. Grav.}, 27:155004, 2010 arXiv:0912.2740,  [gr-qc].

\bibitem{couch:1973}
W.~E. Couch and E.~T. Newman.
\newblock Algebraically special perturbations of the schwarzschild metric.
\newblock {\em J. Math. Phys. (N.Y.), v. 14, no. 2, pp. 285-286}, 2 1973.

\bibitem{Dolan:2009nk}
S.~R. Dolan and A.~C. Ottewill.
\newblock {On an Expansion Method for Black Hole Quasinormal Modes and Regge
  Poles}.
\newblock {\em Class. Quant. Grav.}, 26:225003, 2009 arXiv:0908.0329,  [gr-qc].

\bibitem{1984PhRvD..30..295F}
V.~{Ferrari} and B.~{Mashhoon}.
\newblock {New approach to the quasinormal modes of a black hole}.
\newblock {\em Phys. Rev. D}, 30(2):295--304, July 1984.

\bibitem{gomez1997}
R Gómez, L Lehner, P Papadopoulos and J Winicour
\newblock The eth formalism in numerical relativity.
\newblock {\em Classical and Quantum Gravity}, 14(4):977, 1997.


\bibitem{Kokkotas:1999bd}
K.~D. Kokkotas and B.~G. Schmidt.
\newblock {Quasinormal modes of stars and black holes}.
\newblock {\em Living Rev. Rel.}, 2:2, 1999 arXiv:gr-qc/9909058.

\bibitem{Konoplya:2011qq}
R.~A. Konoplya and A.~Zhidenko.
\newblock {Quasinormal modes of black holes: From astrophysics to string
  theory}.
\newblock {\em Rev. Mod. Phys.}, 83:793--836, 2011 arXiv:1102.4014,  [gr-qc].

\bibitem{1985RSPSA.402..285L}
E.~W. {Leaver}.
\newblock {An Analytic Representation for the Quasi-Normal Modes of Kerr Black
  Holes}.
\newblock {\em Proceedings of the Royal Society of London Series A},
  402(1823):285--298, December 1985.

\bibitem{1996CQGra..13..233L}
H.~{Liu} and B.~{Mashhoon}.
\newblock {On the spectrum of oscillations of a Schwarzschild black hole}.
\newblock {\em Classical and Quantum Gravity}, 13(2):233--251, February 1996.

\bibitem{cedeno:2015ucy}
C.~E. Cede\~no M. and J.~C.~N. de~Araujo.
\newblock {Master equation solutions in the linear regime of characteristic
  formulation of general relativity}.
\newblock {\em Phys. Rev.}, D92:124015, 2015 arXiv:1512.02836,  [gr-qc].

\bibitem{MaassenvandenBrink:2000iwh}
A.~Maassen van~den Brink.
\newblock {Analytic treatment of black hole gravitational waves at the
  algebraically special frequency}.
\newblock {\em Phys. Rev. D}, 62:064009, 2000 arXiv:gr-qc/0001032.

\bibitem{PhysRevD.87.104016}
T.~M\"adler.
\newblock Simple, explicitly time-dependent, and regular solutions of the
  linearized vacuum einstein equations in bondi-sachs coordinates.
\newblock {\em Phys. Rev. D}, 87:104016, May 2013.

\bibitem{moncrief:1974}
V.~{Moncrief}.
\newblock {Gravitational perturbations of spherically symmetric systems. I. The
  exterior problem}.
\newblock {\em Annals of Physics}, 88(2):323--342, December 1974.

\bibitem{mongwane:2017}
B.~{Mongwane}.
\newblock {Characteristic formulation for metric f (R ) gravity}.
\newblock {\em Phys. Rev. D}, 96(2):024028, July 2017 arXiv:1707.05757,
  [gr-qc].

\bibitem{Madler:2016xju}
T.~Mädler and J.~Winicour.
\newblock {Bondi-Sachs Formalism}.
\newblock {\em Scholarpedia}, 11:33528, 2016 arXiv:1609.01731,  [gr-qc].

\bibitem{NewmanPenrose1963}
E.~T. {Newman} and R.~Penrose
\newblock{An approach to gravitational radiation by a method of spin coefficients}
\newblock {\em J. Math Phys} 3:998, 1963.

\bibitem{1993PhRvD..47.5253N}
H.-P. {Nollert}.
\newblock {Quasinormal modes of Schwarzschild black holes: The determination of
  quasinormal frequencies with very large imaginary parts}.
\newblock {\em "Phys. Rev."}, D47(12):5253--5258, June 1993.

\bibitem{1999CQGra..16R.159N}
H.-P. {Nollert}.
\newblock {TOPICAL REVIEW: Quasinormal modes: the characteristic `sound' of
  black holes and neutron stars}.
\newblock {\em Classical and Quantum Gravity}, 16(12):R159--R216, December
  1999.

\bibitem{nzioki:2017}
A.~M. Nzioki, R.~Goswami, and P.~K.~S. Dunsby.
\newblock Covariant perturbations of schwarzschild black holes in f(r) gravity.
\newblock {\em International Journal of Modern Physics D}, 26(06):1750048,
  2017https://doi.org/10.1142/S0218271817500481.

\bibitem{press:1971}
W.~H. {Press}.
\newblock {Long Wave Trains of Gravitational Waves from a Vibrating Black
  Hole}.
\newblock {\em The Astrophysical Journal}, 170:L105, December 1971.

\bibitem{1993GReGr..25.1185Q}
G.~{Qi} and B.~F. {Schutz}.
\newblock {Robinson-Trautman equations and Chandrasekhar's special perturbation
  of the Schwarzschild metric}.
\newblock {\em General Relativity and Gravitation}, 25(11):1185--1188, November
  1993.

\bibitem{regge:1957}
T.~{Regge} and J.~A. {Wheeler}.
\newblock {Stability of a Schwarzschild Singularity}.
\newblock {\em Physical Review}, 108(4):1063--1069, November 1957.

\bibitem{Saad:2003vhv}
N.~Saad, R.~L. Hall, and H.~Ciftci.
\newblock {Asymptotic iteration method for eigenvalue problems}.
\newblock {\em J. Phys. A}, 36(47):11807, 2003.

\bibitem{Sachs103}
R.~K. Sachs.
\newblock Gravitational waves in general relativity. viii. waves in
  asymptotically flat space-time.
\newblock {\em Proceedings of the Royal Society of London A: Mathematical,
  Physical and Engineering Sciences}, 270(1340):103--126,
  1962http://rspa.royalsocietypublishing.org/content/270/1340/103.full.pdf.

\bibitem{Shibata:2006nm}
M.~Shibata and K.~Taniguchi.
\newblock {Merger of binary neutron stars to a black hole: disk mass, short
  gamma-ray bursts, and quasinormal mode ringing}.
\newblock {\em Phys. Rev. D}, 73:064027, 2006 arXiv:astro-ph/0603145.

\bibitem{teukolsky:1972}
S.~A. Teukolsky.
\newblock Rotating black holes: Separable wave equations for gravitational and
  electromagnetic perturbations.
\newblock {\em Phys. Rev. Lett.}, 29:1114--1118, Oct 1972.

\bibitem{vanderWalt:2011jt}
P.~J. van~der Walt and N.~T. Bishop.
\newblock {Observational cosmology using characteristic numerical relativity:
  Characteristic formalism on null geodesics}.
\newblock {\em Phys. Rev. D}, 85:044016, 2012 arXiv:1111.6025,  [gr-qc].

\bibitem{vishveshwara:1970}
C.~V. {Vishveshwara}.
\newblock {Scattering of Gravitational Radiation by a Schwarzschild
  Black-hole}.
\newblock {\em Nature}, 227(5261):936--938, August 1970.

\bibitem{lrr:winicour}
J.~Winicour.
\newblock Characteristic evolution and matching.
\newblock {\em Living Reviews in Relativity}, 15(2), 2012.

\bibitem{zerilli:1970}
F.~J. {Zerilli}.
\newblock {Effective Potential for Even-Parity Regge-Wheeler Gravitational
  Perturbation Equations}.
\newblock {\em Phys. Rev. Lett.}, 24(13):737--738, March 1970.

\bibitem{Zlochower03}
Y. Zlochower, R. Gómez, S. Husa, L, Lehner, and J. Winicour.
\newblock {Mode coupling in the nonlinear response of black holes}.
\newblock {\em Phys. Rev. D}, 68:084014, 2003.

\bibitem{lentz:1976}
W.~J. {Lentz}.
\newblock {Generating Bessel functions in Mie scattering calculations using continued fractions}.
\newblock {\em Appl. Opt.}, 15(3):668--671, 1976.


\bibitem{thompson:1986}
I.~J. Thompson and A.~R. Barnett.
\newblock {Coulomb and Bessel functions of complex arguments and order.}.
\newblock {\em Journal of Computational Physics}, 64(2):490--509, 1986.

\bibitem{gautschi:1967}
W. {Gautschi}.
\newblock { Computational Aspects of Three-Term Recurrence Relations}.
\newblock {\em SIAM Review}, vol. 9, no. 1, pp. 24-82, 1967

\bibitem{baber:1935}
W.~G. Baber and H.~R {Hass{\'e}}. 
\newblock {The Two Centre Problem in Wave Mechanics}.
\newblock {\em Proceedings of the Cambridge Philosophical Society}, vol. 31, no. 4, pp. 564, 1935


\bibitem{samuelsson:2007}
L. {Samuelsson}, N. Andersson and A. Maniopoulou.
\newblock {A characteristic approach to the quasi-normal mode problem}.
\newblock {\em Classical and Quantum Gravity}, vol. 24, no 16, pp. 4147-4160, 2007


\bibitem{Siebel:2001dp}
F. Siebel, J. Font and P. Papadopoulos.
\newblock {Scalar field induced oscillations of neutron stars and gravitational collapse}.
\newblock {\em Phys. Rev. D}, vol. 65,  pp. 024021, 2002

\bibitem{bai:2007}
S. {Bai}, Z. {Cao}, X. {Gong}, Y. {Shang}, X. {Wu}, Y.~K and {Lau}, 
\newblock {Light cone structure near null infinity of the Kerr metric}.
\newblock {\em Phys. Rev. D}, vol. 75, no 4, pp. 044003, 2007

\bibitem{arganaraz:2021}
M.~A. {Arga{\~n}araz}, and O.~M. {Moreschi}
\newblock { Double null coordinates for Kerr spacetime}.
\newblock {\em Phys. Rev. D}, vol. 104, no 2, pp. 024049, 2021

\bibitem{maedler:2023}
T. {M{\"a}dler} and E. {Gallo}
\newblock {Slowly rotating Kerr metric derived from the Einstein equations in affine-null coordinates}.
\newblock {\em Phys. Rev. D}, vol. 107, no 10, pp. 104010, 2023

\bibitem{wald:1979}
R.~M. Wald, 
\newblock {Note on the stability of the Schwarzschild metric}.
\newblock {\em Journal of Mathematical Physics}, vol. 20, no. 6, pp. 1056–1058, 1979.


\bibitem{kay:1987}
B.~S. Kay and R.~M. Wald, 
\newblock {Linear stability of Schwarzschild under perturbations which are nonvanishing on the bifurcation 2-sphere}.
\newblock {\em Journal of Mathematical Physics}, vol. 28, no. 11, pp. 2882–2898, 1987.











\end{thebibliography}
\end{document}